\documentclass[nofootinbib,amsmath,amssymb,pra,11pt,notitlepage]{revtex4-1}

\usepackage{graphicx}			% Include figure files
\usepackage{dcolumn}			% Align table columns on decimalpoint
\usepackage{bm}					% bold math
\usepackage{hyperref}			% add hypertext capabilities
\usepackage{amsmath}
\usepackage{amsfonts}
\usepackage{amssymb}
\usepackage{pstcol,times,color,graphicx,graphics,pstricks,pst-node,pst-coil,pst-grad}
\newcommand{\be}{\begin{equation}}
\newcommand{\ee}{\end{equation}}
\newcommand{\bea}{\begin{eqnarray}}
\newcommand{\eea}{\end{eqnarray}}

\newcommand{\csch}{\textrm{\,csch\,}}

\begin{document}

\preprint{PREPRINT VERSION 1}

\title{Repulsive van der Waals interaction between a quantum particle and a conducting toroid} 

% % % % % % % % % % % % % % % % % % % % % %

\author{P. P. Abrantes}
\email{patricia@if.ufrj.br}
\affiliation{Instituto de F\'{\i}sica, Universidade Federal do Rio de Janeiro \\
 	Avenida Athos da Silveira Ramos, 149, Centro de Tecnologia, Bloco A, Cidade Universit\'aria, Rio de Janeiro-RJ, Brazil}

\author{Yuri Fran\c ca}
\email{yuridiasf@gmail.com}
\affiliation{Instituto de F\'{\i}sica, Universidade Federal do Rio de Janeiro \\
 	Avenida Athos da Silveira Ramos, 149, Centro de Tecnologia, Bloco A, Cidade Universit\'aria, Rio de Janeiro-RJ, Brazil}
	
\author{Reinaldo de Melo e Souza}
\email{reinaldo@if.uff.br}
\affiliation{Instituto de F\'{\i}sica, Universidade Federal Fluminense \\
	Avenida Litor\^anea, s/n, Boa Viagem, Niter\'oi-RJ, Brazil}	
	
\author{F. S. S. da Rosa}
\email{siqueira79@gmail.com}
\affiliation{Instituto de F\'{\i}sica, Universidade Federal do Rio de Janeiro \\
Avenida Athos da Silveira Ramos, 149, Centro de Tecnologia, Bloco A, Cidade Universit\'aria, Rio de Janeiro-RJ, Brazil}

\author{C. Farina}
\email{farina@if.ufrj.br}
\affiliation{Instituto de F\'{\i}sica, Universidade Federal do Rio de Janeiro \\
Avenida Athos da Silveira Ramos, 149, Centro de Tecnologia, Bloco A, Cidade Universit\'aria, Rio de Janeiro-RJ, Brazil}

% % % % % % % % % % % % % % % % % % % % % % 

\date{\today}

\begin{abstract}

We calculate the non-retarded dispersion force exerted on an electrically polarizable quantum particle by a perfectly conducting toroid, which is one of the most common objects exhibiting a non-trivial topology. We employ a convenient method developed by Eberlein and Zietal that essentially relates the quantum problem of calculating dispersion forces between a quantum particle and a perfectly conducting surface of arbitrary shape to a corresponding  classical problem of electrostatics. Considering the quantum particle in the symmetry axis of the toroid,  we use this method  to find an exact analytical result for the van der Waals interaction between the quantum particle and the conducting toroid. Surprisingly, we show that for appropriate values of the two radii of the toroid the dispersive force on the quantum particle is repulsive. This is a remarkable result since repulsion in dispersive interactions involving only electric objects (and particles) in vacuum is rarely reported in the literature. Final comments are made about particular limiting cases as for instance the quantum particle-nanoring system.

\end{abstract}

\maketitle

\pagebreak

%----------------------------------------------------------------------------------------------------------------------------------------

\section{\label{SecIntroduction}Introduction}

Since the advent of quantum mechanics it has been realized that charge, current and field fluctuations  play a crucial role in many phenomena from nano to macroscopic scale. Among these, we may highlight the so-called dispersion forces, which are a direct consequence of those quantum fluctuations and explain quite satisfactorily  the interaction between two neutral and non-polar,  albeit polarizable, molecules. These forces are responsible for different phenomena, varying from the adhesion of geckos to walls \cite{Autumn-2002}, to the stability of colloids \cite{Israelachvili-2011} and the Casimir effect \cite{Casimir-Polder-1946,Casimir-Polder-1948} (for more details on the Casimir effect see, for instance, Ref(s) \cite{Milton-Book-2004,BordagEtAl-Book-2009,Dalvit-Milonni-Roberts-Rosa-2011}; for a detailed discussion on dispersion interactions see Ref. \cite{Buhmann-Book} and references therein, and for a discussion of a variety of quantum vacuum effects see Ref. \cite{Milonni-Book-1994}).

Since the modern era of experiments on Casimir forces, inaugurated by the torsion pendulum experiment made by Lamoreaux \cite{Lamoreaux-1997} in 1997 and followed by many other ingeneous experiments with different techniques \cite{Mohideen-1998,Chan-2001,Decca-2003,Obrecht-2007,vanZwol-2009,Intravaia-2013}, dispersion forces have attracted the attention of researchers of different, but affine, communities, from quantum field theory and quantum optics to the colloidal systems \cite{BordagEtAl-Book-2009,Dalvit-Milonni-Roberts-Rosa-2011,Buhmann-Book,ReviewWoods}. 

Besides, due to the huge technological advances not only in the creation of new materials but mainly in the miniaturization of electromechanical machines, it is mandatory to have a deeper understanding of dispersion forces in the micro to nanoscale for many reasons, namely: for appropriate designs and  satisfactory operations of such devices, as for instance in the development of nano electromechanical  contact switches \cite{Loh-Espinosa-2012} or in the study of the undesired effects of stiction and non-linear behaviour \cite{ChanEtAl-2001,Rodrigues-2018}, to mention just a few. The ultimate idea behind the study of dispersion forces is to control somehow this kind of interaction so that manipulation of atoms, molecules and nanoparticles can be achieved.

Repulsive Casimir forces in dispersive media have been predicted in 1961 by Dzyaloshinskii, Lifshitz and Pitaevskii \cite{DLP-1961}, in a situation where the system is constituted by three different non-magnetic media, namely, two semi-infinite homogeneous dielectrics of permittivities $\epsilon_1(\omega)$ and $\epsilon_2(\omega)$ separated by an infinite slab of a third dispersive homogeneous medium of permittivity $\epsilon_3(\omega)$ such that $\epsilon_1(i\xi) < \epsilon_3(i\xi) < \epsilon_2(i\xi)$ for most frequencies. Repulsive interactions of this kind had not been observed until recently, when Milling and collaborators measured repulsive van der Waals forces with an Atomic Force Microscope \cite{Milling-1996}. In this context, there were many predictions of repulsive dispersive forces in the last decade \cite{RodriguezEtAl-2008,MundayEtAl-2009,Rahi-Zaheer-2010,RodriguezEtAl-2010}.

Another possible route for repulsive Casimir forces involves considering  dielectric-magnetic materials  \cite{Boyer-1974,Cougo-PintoEtAl-1999,KlichEtAl-2002}. For the case of two atoms,  repulsive dispersion forces can appear if one of them  is electrically polarizable while  the other is magnetically polarizable, as shown by Feinberg and Sucher half a century ago \cite{Feinberg-Sucher-1968,Feinberg-Sucher-1970}. This result has also been discussed by  Boyer in 1974 in the context of stochastic QED \cite{Boyer-1974} and by other authors in recent years \cite{Farina-Santos-Tort-2002-AJP,Farina-Santos-Tort-2002-JPA,Buhmann-Book}. Attempts using magnetic metamaterials have already been made \cite{Rosa-Dalvit-Milonni-2008,Pirozhenko-2008,Yannopapas-2009,McCauley-2011} but it turned out that repulsion in such setup  is exceedingly difficult \cite{Rahi-2010}.

There is still another possibility for repulsive dispersion forces  involving non-magnetic bodies which consists in exploring non-trivial geometries. One interesting example of this situation has been reported a few years ago in the literature, namely, the interaction between a small metallic object and an infinitely conducting plate with a circular hole. As shown by numerical methods in Ref(s) \cite{LevinEtAl-2010,McCauleyEtAl-2011}, repulsion can appear if the object lies in the symmetry axis of the circular hole and is electrically polarizable preponderantly in the axis direction. This can be achieved with a needle-like object oriented along the symmetry axis of the hole.  This surprising result was also investigated by many other authors who found analytically that, under similar conditions,  the non-retarded dispersion force between an atom and  a conducting plane with a circular hole is  repulsive for distances from the atom to the center of the hole smaller than $\sim 0.7R$, with $R$ being the radius of the hole \cite{Eberlein-Zietal-2011,ReinaldoEtAl-2011,MiltonEtAl-2011,MiltonEtAl-2012}.

In this work, in order to find other physical situations where repulsive dispersion forces may arise, we  look at a system that, besides possessing a non-trivial geometry,  also involves a non-trivial topology.  With this  motivation in mind, we consider a quantum particle near the simplest conducting surface that already exhibits a non-trivial topology, namely, a conducting toroid. In order to obtain a more treatable analytical solution for such a problem, we compute the non-retarded dispersion force (van der Waals force) between a quantum particle and a perfectly conducting toroid with the quantum particle lying in the symmetry axis of the toroid. Though the perfectly conducting hypothesis and short-distance regime are conflicting assumptions, this choice was made for the following reasons: {\it (i)} it will work for molecules whose dominant transition wavelengths allow the existence of a window of distances to the conducting surface  that are far enough so that the surface can be considered in a first approximation as perfectly conducting but not too far away so that the retardation effects can be neglected; {\it (ii)} we expect that some important features of the interaction, like the attractive/repulsive character of the force will be essentially the same, as it occurs in the atom-plane with a hole system (this can be checked by comparing qualitatively the analytical results obtained in Ref(s) \cite{Eberlein-Zietal-2011,ReinaldoEtAl-2011,MiltonEtAl-2011,MiltonEtAl-2012} with the exact numerical solution presented in \cite{LevinEtAl-2010}). Indeed, we show that for appropriate values of the two radii of the toroid the dispersive force on the quantum particle is repulsive. 					                     

This paper is organized as follows: in the next section we introduce the toroidal coordinates and establish the basic electrostatic results to be used later. In section III we use the Eberlein-Zietal method  \cite{Eberlein-Zietal-2007} to obtain an exact analytical result for the van der Waals interaction in the quantum particle-toroid system which, as already mentioned, can be repulsive for appropriate choices of the system parameters. Section IV is left for final comments and conclusions.

%----------------------------------------------------------------------------------------------------------------------------------------

\section{\label{SecToroidal} Preliminary electrostatic results}

In this section we briefly introduce the toroidal coordinates and use them to discuss the electrostatic problem of a point charge near a grounded conducting toroid. As we shall see in section III, the quantum particle-toroid dispersion interaction is closely related to the interaction of a point charge with the toroid. 

%-----------------------------------------------------------------

\subsection{Toroidal coordinates}

Toroidal coordinates $(\xi, \eta, \phi)$ are usually defined in terms of the cartesian ones $(x, y, z)$ by \cite{Lebedev-Book}
\bea
	x = \frac{f \sinh \xi \cos \phi}{\cosh \xi - \cos \eta} \,, \;\;\; y = \frac{f \sinh \xi \sin \phi}{\cosh \xi - \cos \eta} \,, \;\;\; z = \frac{f \sin \eta}{\cosh \xi - \cos \eta} \,,
\eea

\noindent where the range of the these coordinates are $0 \leq \xi < \infty$, $ - \pi \leq \eta < \pi$ and $0 \leq \phi < 2\pi$, with $f$ being a constant scale factor. It can be shown that surfaces described by constant values of $\xi$ define toroidal surfaces centered at the origin. For instance, the toroidal surface defined by $\xi = \xi_0$ is given by the equation
\be
	(r - f \coth \xi_0)^2 + z^2 = \left( \frac{f}{\sinh \xi_0} \right)^2 \,,
\ee

\noindent where $r = \sqrt{x^2 + y^2}$. The two radii associated with this surface are $a := f \coth \xi_0$ and $b := f \csch \xi_0$, the former being the radius from the  geometrical center of the toroid to any point located at the center of the torus tube and the latter being the radius of the torus tube. From the previous  expressions for $a$ and $b$ it follows immediately that 
\be
	f = \sqrt{a^2 - b^2} \;\;\;\;\; \textrm{and} \;\;\;\;\; \cosh \xi_0 = \frac{a}{b} \,.
\ee

\noindent It can also be shown that surfaces of constant values of $\eta$ are given by spherical calottes whose centers lie along the ${\cal O}z$ axis. A surface defined by $\eta = \eta_0$ is described by the equation
\be
	(z - f \cot \eta_0)^2 + r^2 = \left( \frac{f}{\sin \eta_0} \right)^2 \, .
\ee

\begin{figure}[h!]
	\begin{center}
		\includegraphics[width=7.8cm]{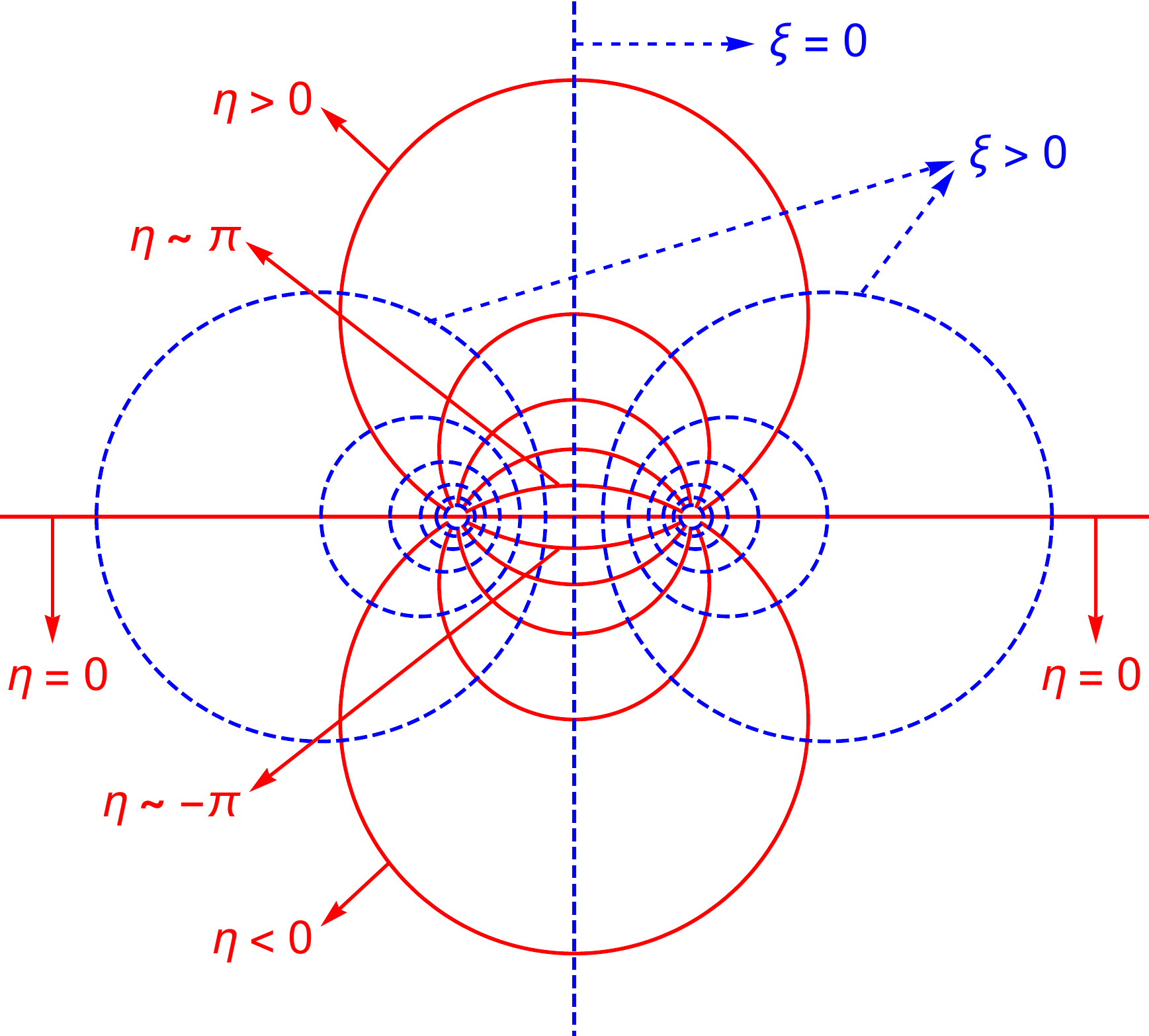}
	\end{center}
\vskip -0.6cm
\caption{(color online) Toroidal surfaces characterized by constant values of $\xi$ (dashed blue lines) and spherical calottes characterized by constant values of $\eta$ (solid red lines).}
\label{SurfacesXiEta}
\end{figure}

\noindent This corresponds to a spherical calotte of radius $f \csc \eta_0$ centered at point $(x, y, z) = (0, 0, f \cot \eta_0)$. The surfaces of constant $\phi$ define planes that contain the ${\cal O}z$ axis. Figure \ref{SurfacesXiEta} depicts sections of different surfaces of constant $\xi$ and $\eta$ for a fixed $\phi$. The limiting case where $\xi \rightarrow \infty$ corresponds to the ring described by $r = f$ and $\xi = 0$ corresponds to the ${\cal O}z$ axis. In addition, it is possible to demonstrate that $\eta = 0$ and $\eta = \pi$ determine, respectively, an infinite plane with a circular hole and a disk, both centered at the origin and with radius $f$.

\vskip -0.3cm

%-----------------------------------------------------------------

\subsection{Point charge near a grounded conducting toroid}

\begin{figure}[h!]
	\begin{center}
		\includegraphics[width=12.5cm]{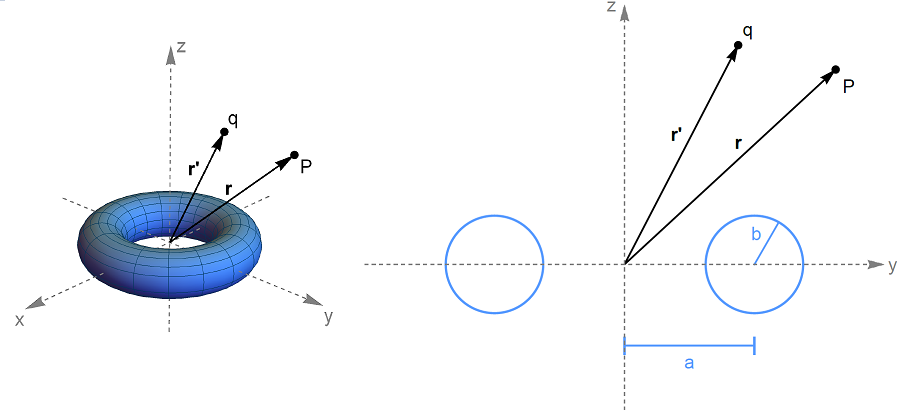}
	\end{center}
\caption{The figure on left shows a generic point of space $P$ at position ${\bf r}$ and a point charge $q$ at position ${\bf r}^\prime$ near a perfectly conducting toroid of radii $a$ and $b$. The figure on the right shows the vertical plane defined by $\phi = \pi/2$ so that the two radii $a$ and $b$ can be easily identified. For convenience, in both figures, the point charge $q$ and the point $P$ belong to this vertical plane.}
\label{PointChargeToroid}
\end{figure}

Let us consider a point charge $q$ at position ${\bf r}^\prime$ near a grounded perfectly conducting toroid defined by $\xi = \xi_0$. Our purpose here is to determine the electrostatic potential created by this system at any point of space outside the toroid and, in particular, to identify the contribution of the superficial charges induced on the toroidal surface to this potential. We start by considering the point charge at an arbitrary point outside the toroid but then, for future convenience, we will assume the point charge to be located at a fixed point of the ${\cal O}z$ axis. This system is shown in figure \ref{PointChargeToroid}. In our notation, primed coordinates refer to the position of the point charge while primeless ones refer to points of space where the electrostatic potential will be calculated.

In order to evaluate the electrostatic potential $V ({\bf r})$ at any point ${\bf r}$ outside the toroid, we must solve Poisson equation, $\nabla^2 V ({\bf r}) = - \mbox{\large$(q/\epsilon_0)$} \delta ({\bf r} - {\bf r}^\prime)$, subject to the appropriate boundary condition (BC) on the toroidal surface $S$. For a grounded toroid, the BC is $V({\bf r}) \Big|_{{\bf r} \in S} = 0$. The desired solution can be written as
\be
	V ({\bf r}) = \frac{q}{4 \pi \epsilon_0} \frac{1}{|{\bf r} - {\bf r}^\prime|} + V_H ({\bf r}) \,,
\ee

\noindent where $V_H ({\bf r})$ satisfies Laplace equation, $\nabla^2 V_H ({\bf r}) = 0$, but subject to the following non-trivial BC
\be
	V_H ({\bf r}) \Big|_{{\bf r} \in S} = - \frac{q}{4 \pi \epsilon_0} \frac{1}{|{\bf r} - {\bf r}^\prime|} \Bigg|_{{\bf r} \in S} \,.
\label{BC1}
\ee

\noindent Although Laplace equation is not immediately separable in toroidal coordinates, separation of variables can be achieved by setting
\be
	V_H (\xi, \eta, \phi) = \sqrt{\cosh \xi - \cos \eta} \, F (\xi) H (\eta) \Phi(\phi) \,.
\ee

\noindent From now on we shall assume the point charge is located on the ${\cal O}z$ axis, so that the problem exhibits an axial symmetry around this axis. A direct consequence of this assumption is that both $V$ and $V_H$ are functions of coordinates $\xi$ and $\eta$, but not $\phi$. Since the solution must be valid for the entire region outside the toroid ($0 \leq \xi \leq \xi_0$), the solution takes the form \cite{Lebedev-Book}
\be
	V_H (\xi, \eta) = \sqrt{\cosh \xi - \cos \eta} \sum_{n = 0}^{\infty} P_{n - 1/2} (\cosh \xi) \Big[ A_n \cos (n \eta) + B_n \sin (n \eta) \Big] \,,
\label{VhGeral}
\ee
where $\left \{ P_\nu(z) \right\}$ are the Legendre functions and coefficients $A_n$ and $B_n$ can be determined by imposing the BC (\ref{BC1}). With this purpose, it is convenient to write $1/|{\bf r} - {\bf r}^\prime|$ in toroidal coordinates. Taking into account the axial symmetry, it can be shown that \cite{Scharstein-Wilson-2005}
\begin{align}
	\frac{1}{|{{\bf r} - {\bf r^\prime}}|} &= \frac{1}{\pi f} \left( \cosh\xi - \cos \eta \right)^{1/2} \left( \cosh \xi' - \cos \eta' \right)^{1/2} \times\nonumber \\
	&\times \sum_{n=0}^{\infty} \left( 2 - \delta_{0n} \right) Q_{n - 1/2} (\cosh \xi_>) P_{n - 1/2} (\cosh \xi_<) \, \cos [n (\eta - \eta')] \, ,
\end{align}

\noindent where $\xi_>$ and $\xi_<$ are the greater and the smaller value between $\xi$ and $\xi^\prime$. Since, by assumption, the point charge is situated on the ${\cal O}z$ axis, we may set $\xi^\prime = 0$. Hence, identifying $\xi_< = \xi^\prime$ and $\xi_> = \xi$, the last equation takes the form
\be\label{Identidade}
\frac{1}{|{{\bf r} - {\bf r^\prime}}|} = \frac{1}{\pi f} \left( \cosh\xi - \cos \eta \right)^{1/2} 
\left( 1 - \cos \eta' \right)^{1/2} \sum_{n=0}^{\infty} \left( 2 - \delta_{0n} \right) Q_{n - 1/2} (\cosh \xi) \cos [n (\eta - \eta')] \,.
\ee

\noindent Substituting  Eq(s) (\ref{VhGeral}) and (\ref{Identidade}) into Eq. (\ref{BC1}) we get
\begin{align}
	\sum_{n = 0}^{\infty} &P_{n - 1/2} (\cosh \xi_0) \Big[ A_n \cos (n \eta) + B_n \sin (n \eta) \Big] =  \nonumber \\
	&- \frac{q}{4 \pi^2 \epsilon_0 f} \left( 1 - \cos \eta' \right)^{1/2} \sum_{n=0}^{\infty} \left( 2 - \delta_{0n} \right) Q_{n - 1/2} (\cosh \xi_0) \left[ \cos (n\eta) \cos (n\eta') + \sin (n \eta) \sin (n \eta') \right] \,.
\label{ApplBC}
\end{align}

\noindent Due to the orthogonality of $\{\sin(n\eta),\, \cos(n\eta);\, n=0,1,2,...\}$, a direct comparison between both sides of Eq. (\ref{ApplBC}) allows us to identify immediately coefficients $A_n$ and $B_n$:
\begin{align}
	A_n &= - \frac{q}{4 \pi^2 \epsilon_0 f} \sqrt{1 - \cos \eta'} \left( 2 - \delta_{0n} \right) \cos (n \eta') \frac{Q_{n - 1/2} (\cosh \xi_0)}{P_{n - 1/2} (\cosh \xi_0)} \,, \\
	B_n &= - \frac{q}{4 \pi^2 \epsilon_0 f} \sqrt{1 - \cos \eta'} \left( 2 - \delta_{0n} \right) \sin (n \eta') \frac{Q_{n - 1/2} (\cosh \xi_0)}{P_{n - 1/2} (\cosh \xi_0)} \,.
\end{align}

\noindent Finally, substituting the previous expressions for $A_n$ and $B_n$ into Eq. (\ref{VhGeral}), we obtain
\begin{align}
	V_H (\xi, \eta, \xi' = 0, \eta'; \xi_0) &= - \frac{q}{4 \pi^2 \epsilon_0 f} \left( \cosh \xi - \cos \eta \right)^{1/2} \left( 1 - \cos \eta' \right)^{1/2} \nonumber \\
	&\times \sum_{n = 0}^{\infty} \left( 2 - \delta_{0n} \right) \cos [n (\eta - \eta')] \frac{Q_{n - 1/2}(\cosh \xi_0) P_{n - 1/2} (\cosh\xi)}{P_{n - 1/2} (\cosh\xi_0)} \,.
\label{VhParticular}
\end{align}

\begin{figure}[h!]
	\begin{center}
		\includegraphics[width=9.0cm]{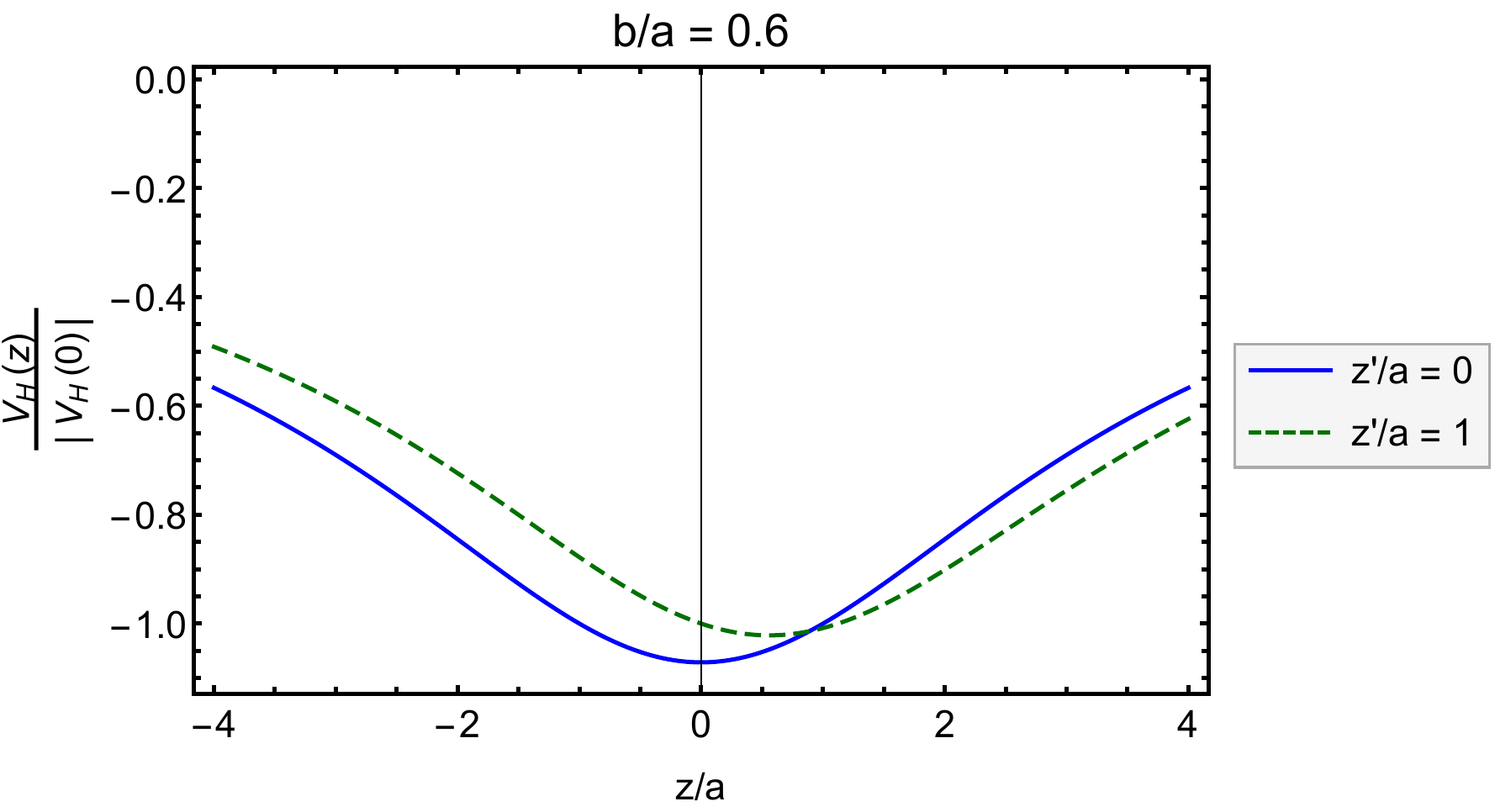}
	\end{center}
\caption{(color online) Electrostatic potential along the ${\cal O}z$ axis, normalized by the absolute value of the electrostatic potential at $z = 0$, created by the superficial charges on the toroidal surface induced by a point charge located at the origin (blue solid line) and located at a point on the positive semiaxis ${\cal O}z$ (green dashed line). In this figure, we chose $a = 5 nm$.}
\label{VH1}
\end{figure}

It is worth emphasizing that the previous expression for $V_H (\xi, \eta, \xi' = 0, \eta'; \xi_0)$, valid for $0\le \xi \le \xi_0$, is precisely the electrostatic potential at any point outside the toroid created by the surface charge distribution induced on the toroidal surface by the presence of the point charge $q$ located on the ${\cal O}z$ axis. For points belonging to the ${\cal O}z$ axis, we just set $\xi = 0$.  In figure \ref{VH1} we plot $V_H (\xi=0, \eta, \xi' = 0, \eta'; \xi_0)$ as a function of $z$ for two different positions of the point charge $q$ (recall that for $\xi = 0$, $\eta$ and $z$ are related by $z = f\mbox{coth}(\eta/2)$). The solid line corresponds to the case where the point charge is placed at the origin whereas the dashed one is for the case in which the point charge is placed above the origin. Note that in the former case, the curve is symmetric with respect to $z = 0$, as expected, since in this case the induced charge distribution on the toroidal surface is symmetric with respect to the ${\cal O}xy$ plane. However, in the latter situation (with the point charge lying above the origin) this is not the case, since this symmetry is lost. A simple way of understanding this result is the following: assume, for instance, that $q>0$. Hence,  there will be more negative induced charges on the upper half of the toroidal surface than in the lower half. This behavior is shown in figure \ref{VH1} (dashed line). Moreover, observe that $V_H (\xi=0, \eta, \xi' = 0, \eta'; \xi_0)$  tends to zero as $z \longrightarrow \pm \infty$, as expected, since it is the electrostatic potential created by the localized charge distribution induced on the toroidal surface.

\begin{figure}[h!]
	\begin{center}
		\includegraphics[width=10.0cm]{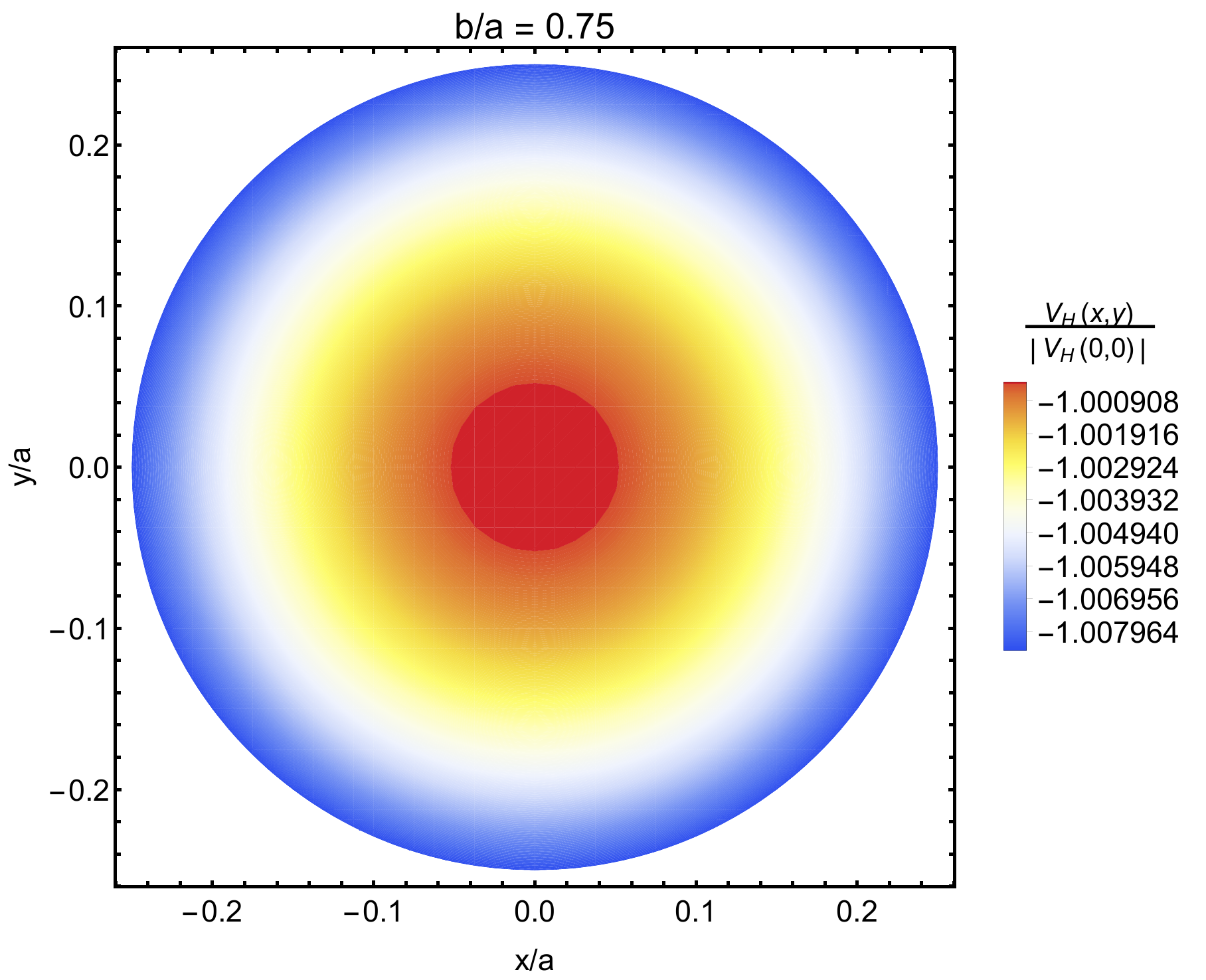}
	\end{center}
\vskip -0.5cm
\caption{(color online) Electrostatic potential, for points on the ${\cal O}xy$ plane outside the toroid but with $0 \le r < a - b$ (which means $\eta=\pi$), created by the superficial charges on the toroidal surface induced by a point charge located at the origin, normalized by the absolute value of the electrostatic potential at $x = y = 0$. In other words, the figure shows $V_H (\xi, \eta=\pi, \xi' = 0, \eta'; \xi_0)$ as a function of $\xi$, which means as a function of $r$, since for $\eta=\pi$ these variables are related by $r = f\tanh(\xi/2)$.  In this figure, we chose $a = 4 nm$.}
\label{VH3}
\end{figure}

The previous graphs depicted in figure \ref{VH1}  showed the behavior, along the ${\cal O}z$ axis, of the electrostatic potential created by the superficial charges on the toroidal surface induced by a point charge located at a point on the ${\cal O}z$ axis. In other words, these graphs showed  $V_H (\xi=0, \eta, \xi' = 0, \eta'; \xi_0)$ as a function of $\eta$ (which means as a function of $z$) for two different values of $\eta^\prime$ (two positions of the point charge). One could as well investigate the behavior of $V_H (\xi, \eta, \xi' = 0, \eta'; \xi_0)$ in other points of the space. For instance, for points on the ${\cal O}xy$ plane outside the toroid but with $0 \le r < a - b$ (which means $\eta=\pi$), with the point charge located at the origin,  $V_H (\xi, \eta=\pi, \xi' = 0, \eta'; \xi_0)$ is shown in figure \ref{VH3}.

\begin{figure}[h!]
	\begin{center}
		\includegraphics[width=9.0cm]{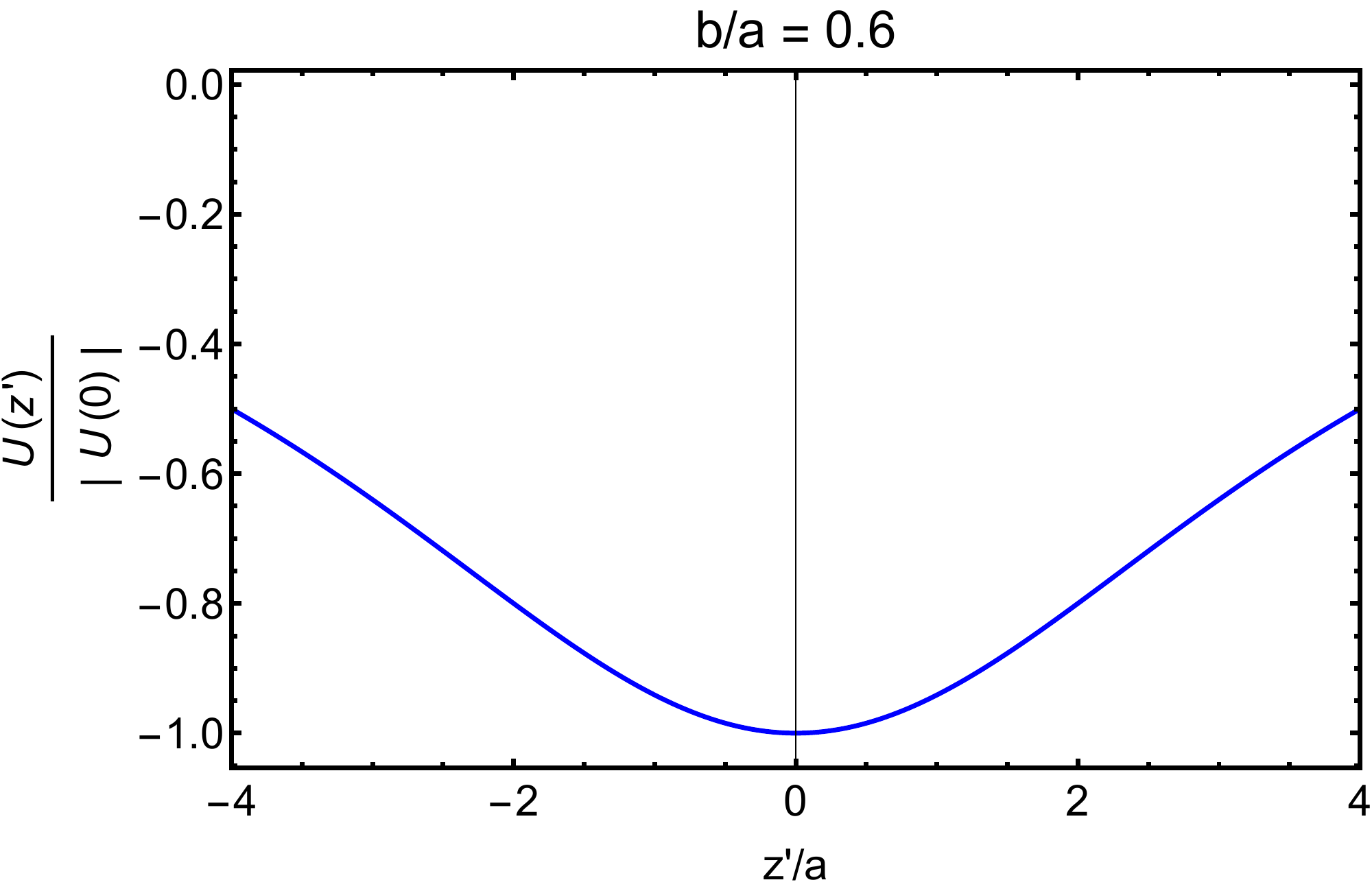}
	\end{center}
\caption{Electrostatic interaction energy, $U(z')$ normalized by $|U (0)|$, between the point charge $q$ at position $(0, 0, z')$ with the induced charges on the toroidal surface. Note that the electrostatic force acting on the point charge $q$ when it lies along the ${\cal O}z$ axis is always attractive, {\it i.e.}, pointing to the origin, except when the charge is at the origin. In this case the electrostatic force vanishes, as expected. In this figure, we chose $a = 5 nm$.}
\label{VH2}
\end{figure}

The previous graphs presented in figures \ref{VH1} and \ref{VH3} show only the values of $V_H$ for different points of space but with the point charge fixed at a given point on the ${\cal O}z$ axis. However, one could be interested in the interaction between the point charge and the charge distribution induced by this point charge on the conducting surface. The electrostatic interaction energy between the point charge $q$ at position ${\bf r}'$ with the induced surface charges is simply given by $qV_H({\bf r}', {\bf r}')$. For the charge-toroid system, the electrostatic energy between the point charge $q$ at position $(0, 0, z')$ and the superficial charge distribution induced by $q$ on the toroidal surface is given by  \footnote{Had we been interested in the total electrostatic energy of the system constituted by the point charge $q$ and the induced superficial charge distribution an extra factor of $1/2$ should be included in Eq. (\ref{UqVH}), to take into account the self-energy of the induced charges on the toroidal surface (see Ref. \cite{Taddei-2009} for a simple discussion of this issue).}
\be\label{UqVH}
U(\eta') = q V_H (\xi=0, \eta=\eta', \xi' = 0, \eta'= \eta'; \xi_0) \,.
\ee
Figure \ref{VH2} shows $U(z')/U(0)$ {\it versus} $z'$ (recall that for points on the ${\cal O}z$  axis, we have $z' = f\mbox{coth}(\eta'/2)$). Concerning motions of the point charge only along the ${\cal O}z$ axis, we see that there is one stable equilibrium point at $z' = 0$, as expected. It is worth emphasizing that this is not a true stable equilibrium point, since this would violate Earnshaw's theorem, which states that it is impossible to have a stable equilibrium point in vacuum only with electrostatic forces. Note, also, that for any other point of the ${\cal O}z$ axis, the charge is attracted towards the origin. 

If, instead of a point charge we consider a point electric dipole oriented along the ${\cal O}z$ axis things will change drastically and repulsive forces may arise depending on the choices of the radii of the toroid. However, instead of analyzing the classical problem of the electrostatic interaction between a point dipole near a grounded perfectly conducting toroid, we shall deal in the next section with the quantum problem, namely, that of a polarizable quantum particle near the grounded conducting toroid. 

%----------------------------------------------------------------------------------------------------------------------------------------

\section{\label{SecNanoring} Repulsion in the quantum particle-toroid system}

\begin{figure}[h!]
	\begin{center}
		\includegraphics[width=6.0cm]{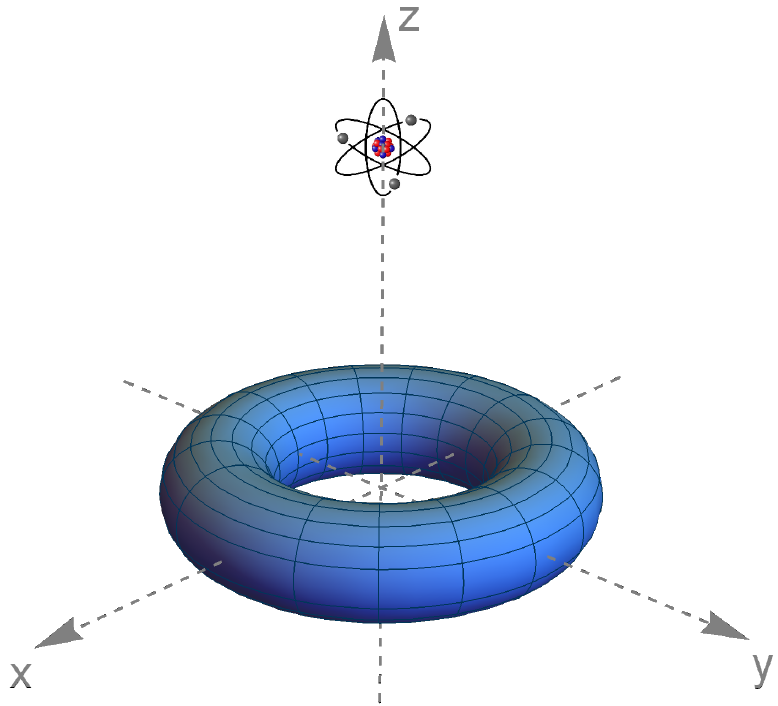}
	\end{center}
\vskip -0.5cm
\caption{Quantum particle at an arbitrary position on the symmetry axis (chosen as the ${\cal O}z$ axis) of a grounded perfectly conducting toroid of radii $a$ and $b$.}
\label{AtomToroid}
\end{figure}

In this section, we shall be concerned with the non-retarded dispersive interaction between an quantum particle and a grounded perfectly conducting toroid. For our purposes, we shall consider the quantum particle fixed at an arbitrary point along the symmetry axis of the toroid, the ${\cal O}z$ axis, as sketched in figure \ref{AtomToroid}. Since, by assumption, retardation effects are being neglected, quantization of the radiation field is not required.

In order to calculate the van der Waals interaction between the quantum particle and the conducting toroid, we shall apply a very convenient method proposed by Eberlein and Zietal \cite{Eberlein-Zietal-2007} in 2007. In fact, this method enables the evaluation of the non-retarded dispersive interaction between an atom and a perfectly conducting surface of an arbitrary shape by solving a correspondent electrostatic problem, namely, that of a point charge at the same position of the atom near the grounded conducting surface. Since it was introduced \cite{Eberlein-Zietal-2007}, this method has been applied with success in many systems \cite{Contreras-Eberlein-2009,Eberlein-Zietal-2011,ReinaldoEtAl-2011,ReinaldoEtAl-Proceeding-2012,ReinaldoEtAl-2013} and has been generalized to describe  two atoms near a conducting body \cite{Reinaldo-2015}.

Following \cite{Eberlein-Zietal-2007}, it is possible to show that the non-retarded interaction energy between one atom at position ${\bf r}_{p}$ and any grounded perfectly conducting surface is given by
\begin{equation}
	U_{NR} ({\bf r}_{p}) = \frac{1}{2 \epsilon_0} \sum_{m=1}^{3} \langle d_m^2 \rangle \nabla_m \nabla_m^\prime G_H({\bf r}, {\bf r}^\prime)\bigg|_{{\bf r} = {\bf r}^\prime = {\bf r}_{p}} \,,
\label{UNR}
\end{equation}

\noindent where ${\bf d}$ is the atomic dipole operator and $G_H ({\bf r}, {\bf r}^\prime)$ satisfies the Laplace equation, $\nabla^2 G_H ({\bf r}, {\bf r}^\prime) = 0$, submitted to the BC
\begin{equation}
	 G_H ({\bf r}, {\bf r}^\prime) \Big|_{{\bf r} \in S} = - \frac{1}{4 \pi |{\bf r} - {\bf r}^\prime|} \Bigg|_{{\bf r} \in S} \,.
\label{BC2}
\end{equation}
All the information about the geometry of the conductor  is encoded in $G_H ({\bf r}, {\bf r}^\prime)$.

It is important to highlight that, except for multiplicative constants, the equation and the BC satisfied by $G_H$ are the same as those satisfied by the electrostatic potential $V_H$ from the previous section. Therefore, by comparing the two functions, we immediately  identify $G_H  = \epsilon_0 V_H/q$.

The next step is to insert this expression in formula (\ref{UNR}) to describe the non-retarded interaction between the quantum particle and the toroidal surface. This leads to
\be
	U_{NR} (\xi_{p}, \eta_{p}) = \frac{1}{2 \epsilon_0} \left\{ \langle d_\xi^2 \rangle h_\xi h_{\xi^\prime}
	\frac{\partial^2}{\partial\xi \partial{\xi^\prime}} + \langle d_\eta^2 \rangle h_\eta h_{\eta^\prime} 
	\frac{\partial^2}{\partial\eta \partial{\eta^\prime}} \right\} G_H (\xi, \eta, \xi', \eta'; \xi_0) \Bigg|_{\xi = \xi^\prime = \xi_{p};\, \eta = \eta^\prime = \eta_{p}} \,,
\ee

\noindent where $h_\xi$ and $h_\eta$ are the so-called metric coefficients, given in the toroidal case by
\be
	h_\xi = h_\eta = \frac{f}{\cosh \xi - \cos \eta} \,.
\ee

For our purposes, we shall assume that the quantum particle is predominantly polarizable in the direction of the ${\cal O}z$ axis, the symmetry axis of the toroid. This means that we can set $\langle d_x^2 \rangle = \langle d_y^2 \rangle \approx 0$, so that only derivatives with respect to $z$ and $z^\prime$ will be necessary. 
The great motivation for this kind of analysis relies on the fact that repulsive van der Waals forces in the system composed of a quantum particle and an infinite plane with a hole occur when the particle is preferably polarizable in the direction of the symmetry axis of the hole. For an isotropic atom, on the other hand, the repulsion is not observed \cite{Eberlein-Zietal-2011}. 

As we are interested only in the $z$-component of the force, we set $\xi = \xi^\prime = 0$ and, hence, only derivatives with respect to $\eta$ and $\eta^\prime$ are necessary. However, since the force on the quantum particle  is on the ${\cal O}z$ direction it is convenient to make a change of variables to cylindrical coordinates. Doing so, the van der Waals dispersive interaction  between the quantum particle and the toroid is given by
\begin{align}
	U_{NR} (z_{p}) &= - \frac{\langle d^2_ z\rangle}{8 \pi^2 \epsilon_0 f} \frac{\partial^2}{\partial z\partial z^\prime} \left\{ \sqrt{\left\{ 1 - \cos \left[ 2 \cot^{-1} \left( \frac{z}{f} \right) \right] \right\} \left\{ 1 - \cos \left[ 2 \cot^{-1} \left( \frac{z'}{f} \right) \right] \right\}} \right. \nonumber \\
	&\times \left. \sum_{n = 0}^{\infty} \left( 2 - \delta_{0n} \right) \cos \left\{ 2n \left[ \cot^{-1} \left( \frac{z}{f} \right) - \cot^{-1} \left( \frac{z'}{f} \right) \right] \right\} \frac{Q_{n - 1/2} (a/b)}{P_{n - 1/2} (a/b)} \right\} \Bigg|_{z = z' = z_{p}} \nonumber \\
	&= -\frac{\langle d^2_ z\rangle f^3}{4 \pi^2 \epsilon_0 (f^2 + z_{p}^2)^3 \sqrt{\left\{ \cos \left[ 2 \cot^{-1} \left( z_{p}/f \right) \right] -1 \right\}^2}} \nonumber \\
	&\times \sum_{n = 0}^{\infty} \left( 2 - \delta_{0n} \right) \left\{ 4n^2 +1 - (4n^2 - 1) \cos \left[2 \cot^{-1} \left(\frac{z_{p}}{f} \right) \right] \right\} \frac{Q_{n - 1/2} (a/b)}{P_{n - 1/2} (a/b)} \,.
\end{align}

\begin{figure}[h!]
	\begin{center}
		\includegraphics[width=10.0cm]{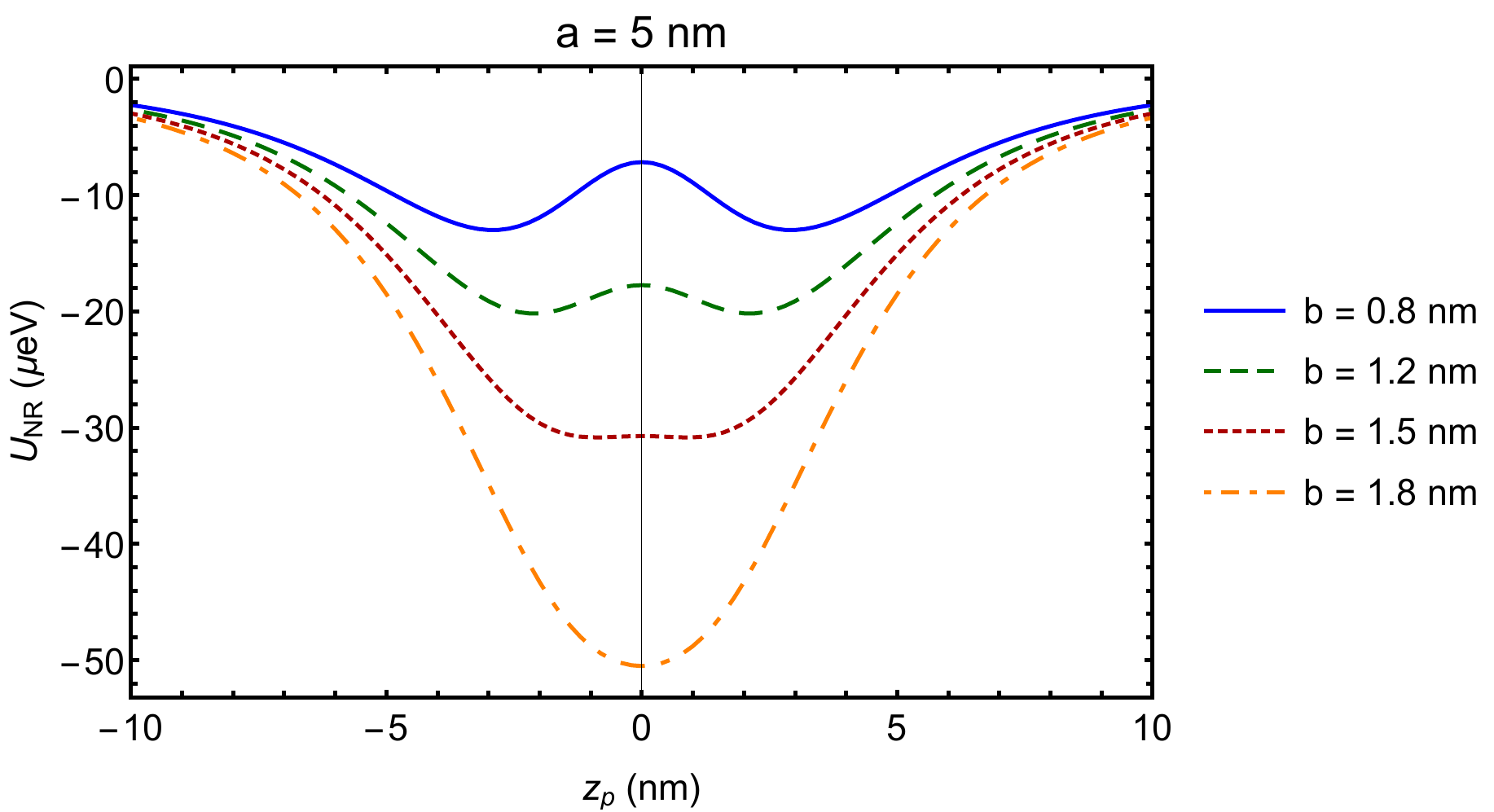}
	\end{center}
\vskip -0.7cm
\caption{(color online) Non-retarded interaction energy $U_{NR}(z_{p})$ as a function of the quantum particle position $z_{p}$, for a fixed value of parameter $a$ and different values of parameter $b$.}
\label{InteractionEnergy}
\end{figure}

We will display our results graphically. The first plot (figure \ref{InteractionEnergy}) shows the non-retarded dispersive energy between the quantum particle and the conducting toroid, $U_{NR}(z_{p})$, as a function of the quantum particle position $z_{p}$, setting a fixed value for the parameter $a$ and varying the values of parameter $b$. A rapid inspection of this graph tells us a quite interesting result, namely, that for small values of $b$ (nanoring limit) the origin is a unstable equilibrium position (only in the ${\cal O}z$ direction) and  the force acting on the quantum particle is repulsive for short distances from the origin. Then, as $z_{p}$ increases, the force on the quantum particle tends to zero and then changes its sign, becoming an attractive force, which goes monotonically to zero as the quantum particle goes to infinity. Note that all curves in this figure are symmetric with respect to the origin. 

\begin{figure}[h!]
	\begin{center}
		\includegraphics[width=10.0cm]{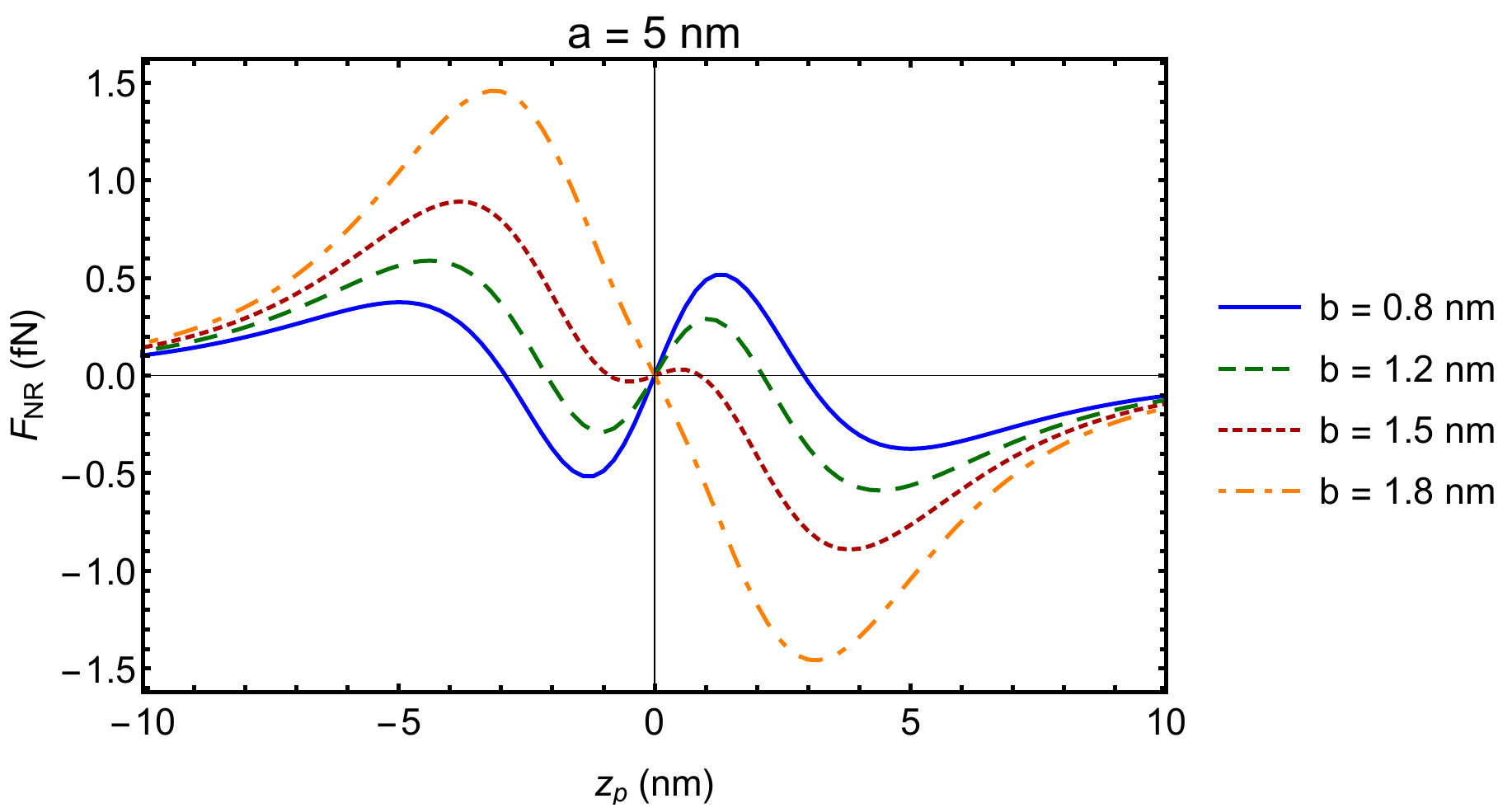}
	\end{center}
\vskip -0.7cm
\caption{(color online) van der Waals dispersive force acting on the quantum particle as a function of the its position $z_{p}$, for a fixed value of radius $a$ and different values of radius $b$. The interval of distances analyzed is the same as that used in the previous figure.}
\label{NRForce1}
\end{figure}

The previous mentioned characteristics become even clearer if we look directly at figure \ref{NRForce1} where the non-dispersive force acting on the quantum particle is represented as a function of its distance $z_{p}$ to the origin for the same values of radii $a$ and $b$ that define the toroidal surface. These curves are obtained essentially just by taking one more derivative of the function $G_H$ with respect to the quantum particle position. Note that all curves in figure \ref{NRForce1} are odd functions of $z_{p}$, as expected. Note, also, that for values of $b$ greater than a given critical value repulsive forces disappear.

\begin{figure}[h!]
	\begin{center}
		\includegraphics[width=10.0cm]{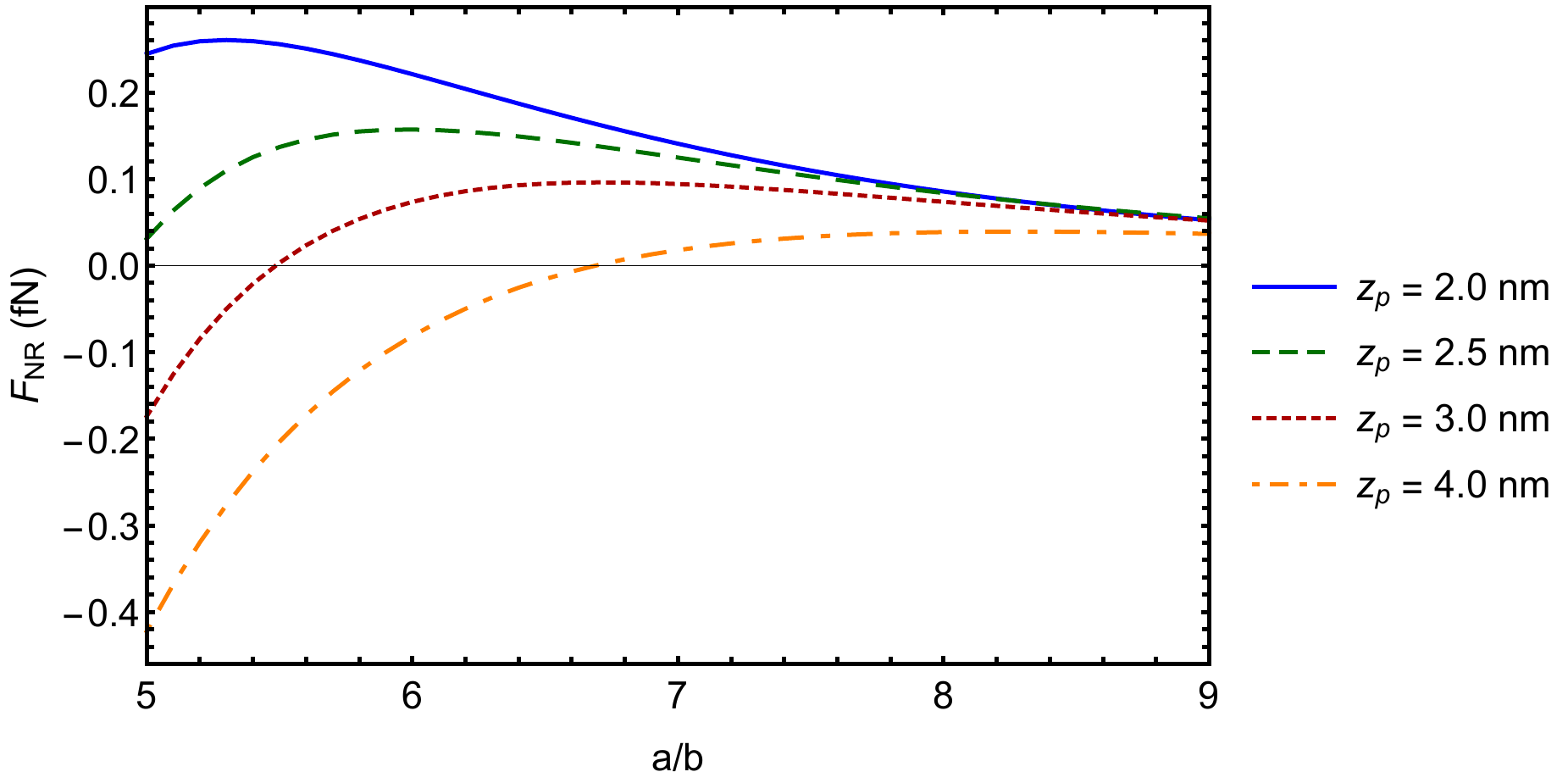}
	\end{center}
\vskip -0.7cm
\caption{(color online) Non-retarded dispersive force as a function of $a/b$ ratio, setting $b = 1 nm$.}
\label{NRForce3}
\end{figure}

In figure \ref{NRForce3}, we choose $b = 1 nm$ and plot the van der Waals force acting on the quantum particle but now as a function of the ratio $a/b$, for different values of the distance from the quantum particle to the origin. It is remarkable to observe that for any distance $z_{p}$, we can always increase the ratio $a/b$ so that the force becomes repulsive. Recall that large values of $a/b$ means the ring limit. Hence, we see that the distance intervals for which repulsive forces occur increase as we approach the ring limit. In other words, as  $z_{p}$ increases we need larger values of the ratio $a/b$ to make repulsive forces possible.

\begin{figure}[h!]
	\begin{center}
		\includegraphics[width=9.5cm]{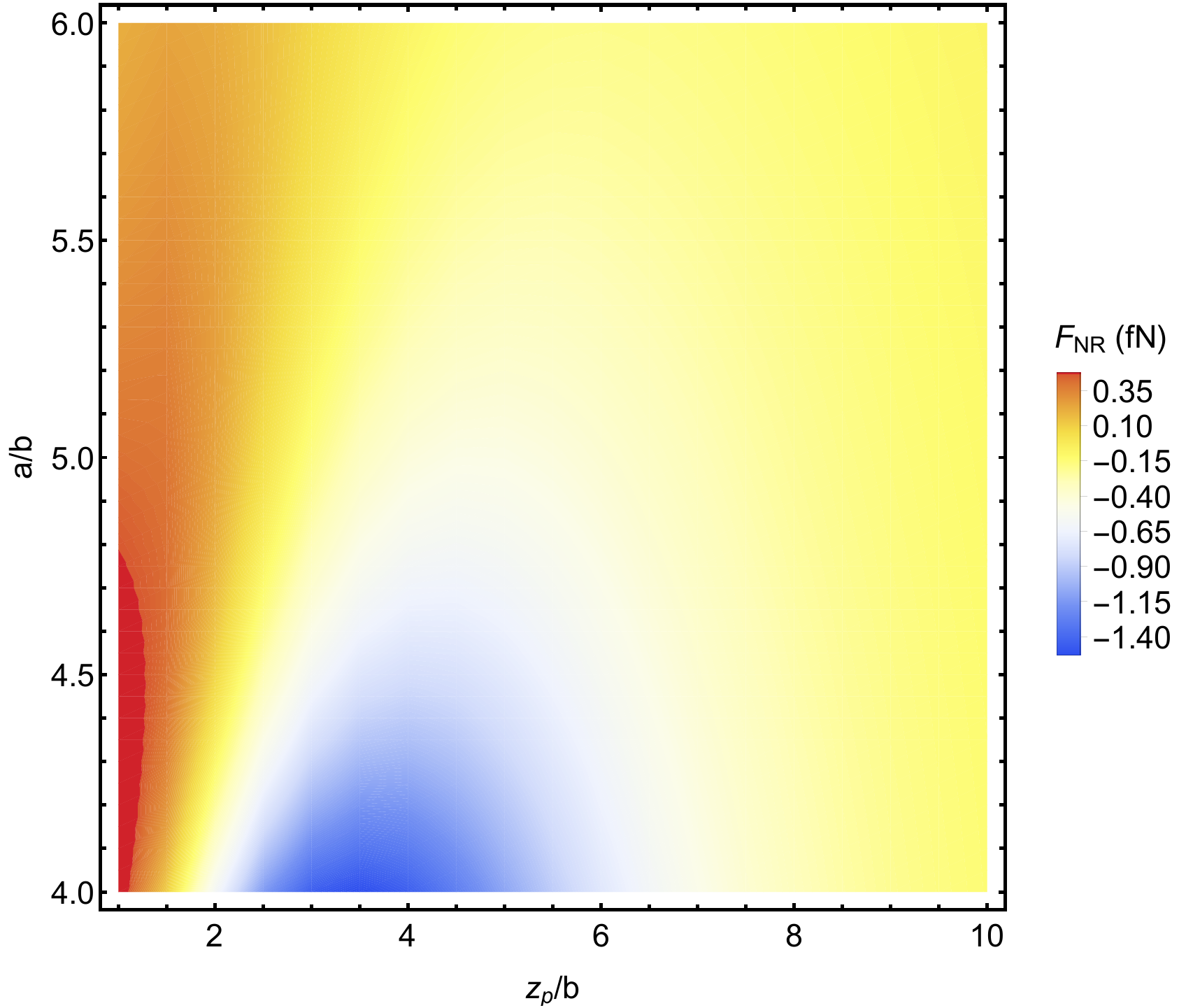}
	\end{center}
\vskip -0.6cm
\caption{(color online) Contour plot depicting the magnitude of the van der Waals force acting on the quantum particle as a function of its distance to the origin, $z_{p}$, and the radius $a$, divided by the radius $b = 1 nm$.}
\label{NRForce4}
\end{figure}

Finally, we have the contour plot shown in figure \ref{NRForce4}, in which the axes are the quantum particle-origin distance $z_{p}$ from the quantum particle to the origin and the radius $a$, both divided by $b$ taken as $1 nm$. The color scale depicts the magnitude of the van der Waals force acting on the quantum particle and it is indicated in the legend on the right. Notice that cuts in this picture, fixing $a/b$ and $z_{p}/b$, are consistent with the results obtained in figure \ref{NRForce1}.

\vskip 0.3cm

%----------------------------------------------------------------------------------------------------------------------------------------

\section{\label{Conclusions} Conclusions and final remarks}

In this work, we have analyzed the non-retarded dispersive forces in a system with a non-trivial topology, namely, a polarizable quantum particle and a perfectly conducting grounded toroid. We solved this problem analitically by using Eberlein-Zietal's method, that is very convenient as all it requires is the solution of an analogous classical problem. For this reason, we presented initially the solution of the electrostatic problem consisting of a point charge in the presence of the same conducting surface.

The most remarkable result was to find another system that presents repulsive and attractive regimes of van der Waals forces, depending on the values ​​of the two radii of the toroid and the distance from the quantum particle to the geometrical center of the toroid. On the one hand, the quantum particle-toroid system exhibits similar features with respect to the well-known example discussed by the MIT group, to wit, a needle-like object near an infinite plane with a circular hole \cite{LevinEtAl-2010}.%, since in both systems the considered objects (an atom in our case and a needle-like object in the MIT case) are only electrically polarizable, the explored geometries are quite non-trivial and the pattern of the van der Waals forces are surprinsingly alike.
On the other hand, we should emphasize an important difference. The qualitative argument given in Ref. \cite{LevinEtAl-2010} to justify the repulsive force acting on the object does not work in the quantum particle-toroid system. Note that while in the former the field lines of a point dipole at the center of the hole cross the infinite plane perpendicularly, the same does not occur with a point dipole at the center of the toroid, which makes this problem somewhat more intriguing.

It is worth mentioning that, at these scales, this type of study is important for controlling and manipulating atom interactions with other systems, since fluctuation-induced forces may be relevant in a variety of situations and can substantially modify the performance of the desired system. A possible application of our results could be in the manufactoring of new designs for atomic mirrors: in general, van der Waals interaction between an atom and a wall is attractive, preventing an efficient reflection by the wall, except for extremely cold atoms, for which quantum reflection may occur \cite{Kouznetsov-2006}. Hence, any mechanism for enhancing repulsive interactions would be welcome. Our work suggests that a wall full of holes, like for instance nano-toroids put side by side forming a toroid regular net would do the job. Maybe this could be an alternative route to the recently ingenious work by Kouznetzov {\it el al}, who made some preliminary estimates showing that profiled ridged surfaces could be used as atomic mirrors.

%----------------------------------------------------------------------------------------------------------------------------------------

\begin{acknowledgments}
The authors are indebted to Daniela Szilard and Yuri Muniz for enlightening discussions. The authors also thank CNPq and FAPERJ for partial financial support.
\end{acknowledgments}

%----------------------------------------------------------------------------------------------------------------------------------------

\end{document}